# Polarization features of optically pumped CdS nanowire lasers

Robert Röder[1], Daniel Ploss[2], Arian Kriesch[2], Robert Buschlinger[2], Sebastian Geburt[1], Ulf Peschel[2] and Carsten Ronning[1]

[1] Friedrich-Schiller-Universität Jena, Institut für Festkörperphysik, 07743 Jena, Germany

[2] Friedrich-Alexander-Universität Erlangen-Nürnberg, Institut für Optik, Information und Photonik, and Erlangen Graduate School in Advanced Optical Technologies (SAOT), Haberstr. 9a, 91058 Erlangen, Germany

Email: robert.roeder@uni-jena.de

**Abstract**. High quality CdS nanowires suspended in air were optically pumped both below and above the lasing threshold. The polarization of the pump laser was varied while the nanowire emission was monitored in a 'head-on' measurement geometry. Highest pump-efficiency and most efficient absorption of the pump radiation are demonstrated for an incident electric field being polarized parallel to the nanowire axis. This polarization dependence was observed both above the lasing threshold and in the regime of amplified spontaneous emission. Measured Stokes parameters of the nanowire emission reveal that due to the onset of lasing the degree of polarization rapidly increases from approximately 15% to 85 %. Both, Stokes parameters and degree of polarization of the nanowire lasing emission are independent of the excitation polarization.

# 1. Introduction

Semiconductor nanowire (NW) photonic lasers are promising to overcome drawbacks of integrated electronic circuits [1, 2] by exploiting their ability as functional connection units between electronic and photonic applications [3]. Recent progress in nanophotonics [4] and plasmonic circuitry [5] emphasizes the need for nanoscale light sources [6]. Besides application of semiconductor nanowires as optoelectronic devices [7, 8, 9], the remarkable optical properties of single nanowires like strong localization of light [10] and efficient waveguiding [11, 12] as well as laser oscillations under pulsed optical pumping were already demonstrated for a wide range of semiconductor materials with emission from the UV to the NIR [13, 14, 15]. Optical pumping of the semiconductor nanowire material enables inversion and high optical gain suitable even for continuous wave lasing [16]. Together with their extremely small size, this makes semiconductor NWs ideal nanoscale coherent light sources for photonic lasing [17], drawing a possible route to optical data transmission and processing. The consequent next step beyond the proof of nanowire lasing is the modification of the emission, for example by adjusting the resonator properties with a focused ion beam machine (FIB) to create coupled cavity NW lasers [18] or loop resonators [19]. Cadmium sulfide (CdS) nanowire lasers bridge the green spectral region around 2.4 eV, provide excellent optical mode confinement [20] and form quasi 1D optical systems with diameters between 150 nm – 500 nm. Optical excitation anisotropy was already observed in a spontaneous excitation regime for ultrathin II-VI semiconductor NWs with diameters below 50 nm [21, 22] and in photoconductivity studies on ZnO semiconductor NWs [23]. Hence, in optical one dimensional CdS NW laser resonators the excitation polarization should influence the absorption cross section and therefore also the pumping efficiencies, which might help to adjust and to modulate the laser emission directed out of the NW laser end facet.

In this study, we present a 'head-on' measurement geometry for the detection of the NW emission characteristics. The nanowires which are partially resting on a substrate and partially suspended in air are optically pumped from above. Simultaneously, the emission from the end facet, which terminates the laser cavity, is directly detected and analyzed. As excitation and detection paths are aligned perpendicularly to each other, they are mutually decoupled and can be controlled separately. We can adjust the polarization of the exciting field relatively to the NW axis, while the emission is analyzed independently for laser oscillations as well as for super luminescence (ASE). Absorption cross sections for parallel and perpendicularly polarized excitation obtained from Finite Difference Time Domain (FDTD) simulations [24] are correlated to the observed difference in pumping efficiency. Our 'head-on' measurement setup is furthermore capable to determine the Stokes parameter $S_0 - S_3$ of the nanowire laser emission and subsequently its degree of polarization.

## 2. Experimental details

Cadmium sulfide nanowires (CdS) were synthesized by a thermal transport technique using the vapour-liquid solid (VLS) mechanism. The growth parameters were chosen according to [14] in order to obtain optically one dimensional, single crystalline CdS NWs. Briefly, CdS powder was evaporated at 800 °C, the Ar carrier gas was supplied at 50 mbar for 30 min to transport the atomic species downstream towards the silicon growth substrate, which was coated beforehand with an Au catalyst. Single NWs were transferred afterwards by a modified dry imprint technique onto the edge of a $SiO_2$/Si substrate (1.5 μm $SiO_2$ layer on top of Si), thus the NWs rest partially on the $SiO_2$/Si substrate and are partially suspended in air, as illustrated in Figure 1(a).

Optical microphotoluminescence (μPL) investigations were performed using a home-built set-up. For the measurement of the excitation anisotropy like in Figure 1(a), the linearly polarized excitation beam of a frequency-tripled Nd:YAG laser (355 nm, 10 ns pulses with a repetition rate of 200 Hz) was focused by a 15x refractive objective (NA = 0.3) to a spot size of approximately 300 μm² on the sample. The polarization of the exciting electric field on the sample could be adjusted with a ratio of at least 1:50. Light emitted from the end facet was collected by a second objective (100x, NA = 0.5) in order to determine the spectral properties of the NW emission, which was dispersed by a 500 mm monochromator and detected by a nitrogen cooled CCD. The Stokes parameter of the NW emission were measured using a high numeric aperture objective (100x, NA = 0.9) for collecting the emission followed by a quarter-wave plate, a linear polarizer and a CCD camera for detection [25].

## 3. Results and discussion

The anisotropy of the pumping efficiency was investigated for the CdS NW shown in the scanning electron microscopy image in Figure 1(b) with a diameter of ~ 500 nm. The power dependent measurements were fitted using a multimode laser model (compare [14]). This demonstrated the lasing ability of the NW with a threshold of approximately 40 kW/cm². The excitation density for the polarization dependent measurement in Figure 1(c) was set to maximum 1.75 times the threshold to avoid destruction of the NW material due to heating, since heat dissipation via the substrate is absent for the part of the NW suspended in air. Although the pumping power was rather close to the threshold, the emission spectra clearly reveal dominant lasing modes superimposed to the nearly negligible spontaneous emission. The lasing mode spacing $\Delta\lambda \approx (1.52 \pm 0.03)$ nm fits the expected value for a 12.5 µm long CdS (n ~ 2.76, dn/dλ ~ - 8.1 µm$^{-1}$ [26]) NW Fabry-Pérot resonator. The µPL spectra in Figure 1(c) display the emission propagating out of the NW end for the polarization of the exciting electric field parallel ($\alpha = 0°$, black), perpendicular ($\alpha = 90°$, red) and diagonal ($\alpha = 45°$, blue) relative to the NW axis. The spectrally broad spontaneous emission, which originates mainly from the excitation at the NW body, remains almost constant and therefore seems to be nearly unaffected by the polarization adjustment. The stimulated emission modes however are highly sensitive to the polarization of the optical pumping beam. They show the highest output intensity for the excitation polarized parallel to the wire axis and the lowest output intensity for perpendicular polarization. The integrated total NW µPL intensity, which is a superposition of the spontaneous and the lasing emission as function of the polarization angle $\alpha$ in Figure 1(d) follows a sin² dependence with highest intensity for parallel adjustment ($\alpha = 0°$) of the exciting electric field revealing a polarization ratio $\rho = (I_\parallel - I_\perp)/(I_\parallel + I_\perp)$ of approximately 0.15. The polarization angle dependence of the stimulated low energy mode A and of the high energy mode B is also displayed in Figure 1(d) and fitted by a sin² function. Both reveal higher polarization ratios $\rho_A = 0.24 \pm 0.06$ and $\rho_B = 0.51 \pm 0.04$ and therefore a rather high sensitivity to changes of the polarization of the exciting electric field.

The response to the parallel and perpendicular adjustment of the excitation polarization was checked for the same CdS NW also in lower ASE regime slightly below lasing threshold, when the spontaneous emission consisting of near band edge (NBE) and waveguide emission still dominates [11], but additionally stimulated modes start to evolve. Figure 2 clarifies the equivalence of the output spectra for both perpendicular polarizations (± 90°), while the more efficient optical pumping of the lasing modes occurs again for the parallel polarization (0°). The spontaneous emission (black data in inset of Fig. 2) is nearly polarization independent, while the stimulated modes (red data) are highly polarization sensitive. This is mainly caused by the low energy mode at ~ 518 nm, which is absent for the perpendicular polarization, but already stimulated for the efficient parallel polarization. The low energy mode B is therefore easily tunable by the pumping polarization below and above threshold, since the gain spectrum is likely to be shifted by higher pumping efficiencies for parallel excitation.

Since the polarization dependence of the lasing output is likely caused by a more efficient supply with carriers to ensure inversion in the semiconductor material for parallel adjusted polarization of the pump beam, the absorption of the 355 nm pumping field was calculated using the FDTD method. Although the CdS NW in Fig. 1(b) exhibits only a small contact area to the substrate, understanding the measurement geometry requires knowledge of the absorption cross sections both of a NW lying on the substrate (open symbols in figure 3) and of one completely surrounded by air (filled symbols). Both values were determined as a function of the diameter of the nanowire and are shown in Figure 3(a). Additionally, the absorption cross section for the free CdS NW was calculated analytically (dashed line). Note, CdS nanowires require diameters above ca. 160 nm for photonic lasing [14]. Below this diameter, the absorption of perpendicularly polarized light is predicted to be larger than for parallel polarization. However, the absorption for pumping with parallel polarization (black data in Fig. 3(a)) starts to exceed that with transversal polarization (red data in Fig. 3(a)) above ca. 190 nm diameter for both configurations, with and without substrate. In general, the influence of the substrate (silica or air) becomes negligible above this diameter. The normalized absorption profiles obtained for 355 nm excitation in the investigated CdS NW are shown in Figures 3(b),(c) for parallel ($\alpha = 0°$) and perpendicular ($\alpha = \pm 90°$) polarization. Absorption profiles almost coincide with and without underlying $SiO_2$ substrate for a NW diameter of 500 nm and have even similar shapes for both polarizations with a high absorption region close to the upper surface and a rather low absorption in the lower part. However, the absolute absorption cross section is increased by a factor of approximately 1.24 for $\alpha = 0°$ compared to $\alpha = \pm 90°$ pumping. Any significant effect of the shape of the absorption profile can therefore be ruled out. Since laser oscillations are likely caused by the formation of an electron-hole plasma (EHP) at high pumping intensities in single CdS (and ZnO NWs as a comparable NW laser system) [14, 16, 27, 17], the approximately 1.24 times higher absorption for parallel polarization supplies the semiconductor more efficiently with charge carriers. This is further accompanied by a red-shift of the gain profile at the same pumping powers due to band gap renormalization effects in the EHP [28]. The red-shift of the gain profile, which in our case is achieved by changing the polarization from $\alpha = \pm\,90°$ to $\alpha = 0°$ and proven by the high polarization ratio for the low energy mode B, is usually obtained for an increase in pumping intensity under constant polarization [14, 17].

By applying modifications, as illustrated in Figure 4, our 'head-on' measurement setup is also suited to determine the Stokes polarization parameters of the NW lasing emission using the rotating quarter-wave plate measurement method [25]. The Stokes polarization parameters $S_0 - S_3$ were determined for a second CdS NW laser with a smaller diameter of ~ 280 nm and a length of 39.9 μm); thus, the parallel pumping is still more efficient according to the calculated absorption cross sections in Figure 3(a). Using 'the rotating quarter-wave plate measurement', the linear polarizer is fixed in vertical position ($\alpha = 0°$) and the emission intensity of the NW laser originating out of the end facet is detected for 8 equidistant positions of the quarter-wave plate in between 0° and 157.5°. The lasing intensity, which is shown as a function of the quarter-wave plate rotation angle in Figure 4 for both excitation polarizations, reveals a 90° periodicity. The 5-7 % fluctuations of the pumping intensity of the Nd:YAG laser system are included in the error bars. By measuring the intensity for eight equidistant data points Nyquist's sampling theorem is fulfilled and the Stokes vectors are determined as: $\vec{S_\parallel} = \begin{pmatrix} 1 \\ -0.78 \\ -0.36 \\ -0.02 \end{pmatrix}$ for parallel (black) and $\vec{S_\perp} = \begin{pmatrix} 1 \\ -0.78 \\ -0.36 \\ 0.05 \end{pmatrix}$ for perpendicular excitation (red) with comparable values. The normalized Stokes parameters are additionally visualized as an intensity plot versus the continuous wave plate rotation angle in Figure 4. The degree of polarization P for the CdS NW lasing is therefore estimated to be 85 ± 15 % and independent of the polarization of the pumping. This value for P represents a lower limit, since the eight data points were acquired individually, thus the polarization in the waveguide NW structure might deviate slightly between the single measurements. Nevertheless, the NW lasing emission contains almost completely linearly polarized light as expected for a laser with a constant transversal laser mode, which is unambiguously the same for both excitation polarizations. Our results, in particular the linear polarization and the high P value, for the CdS nanowire laser emission coincide very well with recent studies on top-down processed GaN nanowires [29]. The degree of polarization P decreases for the low excitation regime to values below 10 – 15 %, where the nanowire emits only spontaneous emission. Hence, a determination of the polarization state of the emitted light allows distinguishing between spontaneous and stimulated emission.

Further, it is possible to characterize the properties of the nanowire guided modes using FDTD simulations. We numerically determined the transverse mode profiles for a nanowire on a $SiO_2$ substrate using the mode solver of the commercial software Lumerical FDTD solutions. We excited a single mode in the nanowire and let it propagate towards the end facet in order to calculate mode reflectivities at the end facet. Using overlap integrals, the reflected field was decomposed into the individual modes [30]. We performed all simulations for a nanowire with a diameter of 280 nm lying on a SiO2 substrate. This is valid especially for the second nanowire, since approximately 90 % of its length was resting on the substrate. Among the four lowest-order modes HE11, TE01, TM01 and HE21, the HE21 mode has the highest reflectivity of 56 %. The remaining modes have reflectivities in the range between 20 % and 30 %. Using the method described in [31], also the confinement factors $\Gamma = \frac{2\varepsilon_0 c n_{wire} \int_{wire} |\vec{E}|^2 dxdy}{\int_\infty Re\{\vec{E}\times\vec{H}^*\}\vec{z} dxdy}$ of the four lowest order transverse modes were determined and plotted in Figure 5 versus the nanowire diameter. As the substrate breaks the symmetry, the two HE11 modes, as well as the TE01 mode and the TM01 mode are no longer degenerate. For small diameters close to the cut-off also the confinement factor is slightly influenced by the substrate and differs from that of a nanowire suspended in air. It turns out, that the experimentally determined lower diameter limit of 170 nm – 180 nm for low threshold lasing [14] coincides with the cut-off of the HE21 mode, indicating that the first mode that starts to lase with increasing diameter is very likely the HE21 mode. To find the most probable lasing mode for the experimental diameter of 280 nm, we calculated the lasing threshold condition: $\Gamma \cdot g_{th} = \frac{1}{L} \cdot ln\frac{1}{R^2}$, where $\Gamma$ is the confinement factor, $R$ the reflectivity assumed to be approximately the same for both end facets and $L$ the wire length [32, 10]. Waveguide losses are neglected. From this, the material gain necessary to achieve lasing can be determined. Using this relation, we again find the HE21 mode to be the most probable lasing mode with lowest required material gain.

## 4. Conclusion

Cadmium sulfide NWs synthesized using the VLS technique were deposited at the edge of the substrate to detect the emission directed out of the NW end facet without scattering from the underlying substrate in a 'head-on' measurement setup. The polarization of the optical pumping beam was systematically adjusted relatively to the NW axis, revealing the most intense output for an excitation polarized parallel to the nanowire axis, a result, which agrees well with respective numerical and analytical simulations. The absorption cross section indicates a higher pumping efficiency for the pump polarization parallel to the nanowire, which is accompanied by a red-shift of the gain profile leading to an increased polarization ratio $\rho$ for low energy modes. By applying modifications to the 'head-on' setup, the degree of polarization P of the CdS NW emission was determined and a significant increase of the polarization degree from 10-15 % for spontaneous emission to $P \approx 85 \pm 15$ % for lasing was observed for both excitation polarizations. Since the Stokes parameters of the NW lasing are also the same for both excitations, we conclude that the transversal lasing mode, which is most likely the HE21 mode, is not notably affected by the polarization of the pumping beam in contrast to the supply with optical gain.

**Acknowledgements**

The authors gratefully acknowledge funding by the German Research Society (DFG) within the project FOR1616. DP and AK acknowledge support by the Erlangen Graduate School in Advanced Optical Technologies (SAOT) and the Cluster of Excellence Engineering of Advanced Materials (EAM), both within the Excellence Initiative of the German Research Foundation (DFG). RB acknowledges support by the International Max Planck Research School Physics of Light.

**Figures**

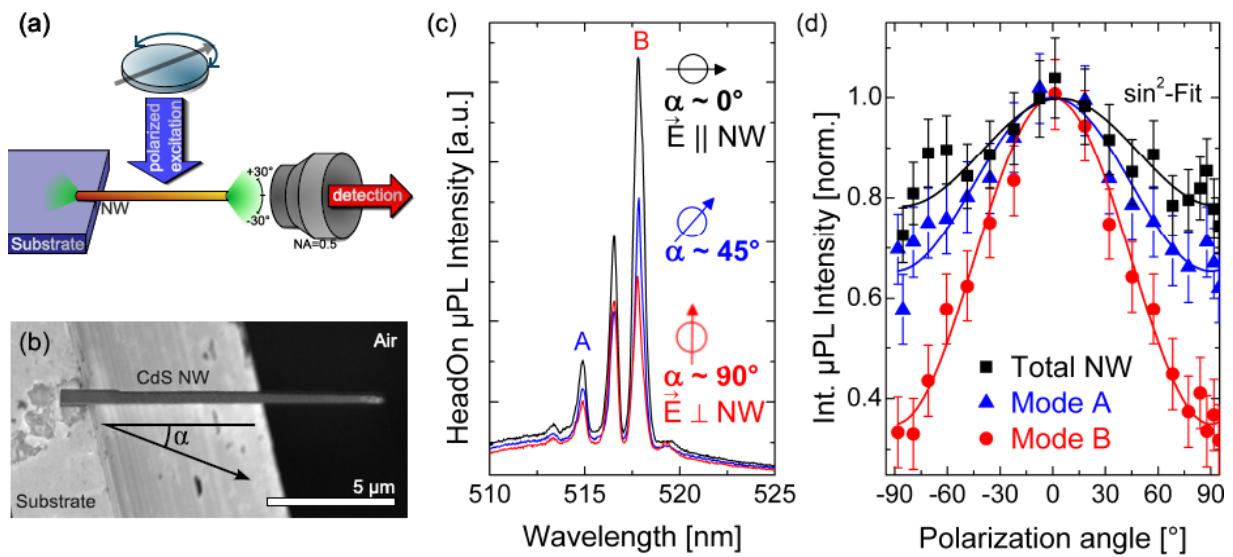

**Figure 1.** (a) Illustration of the 'head-on' measurement geometry with decoupled excitation and detection light path. The CdS NW is partially suspended in air, the excitation polarization is optically adjusted relatively to the NW axis, while the NW laser emission is detected within a cone of 60°. (b) SEM image of the investigated CdS NW (diameter ~ 500 nm, Length = 12.5 μm). (c) RT lasing spectra at an optical pumping of 1.75x the threshold value of approximately 40 kW/cm² for the polarization of the exciting electric field parallel (α = 0°, black), perpendicular (α = 90°, red) and diagonal (α = 45°, blue) relative to NW axis. (d) Polarization dependence of the integrated NW emission (black squares), the intensity of the highest energy mode A (blue triangles) and the lowest energy mode B (red circles) are fitted using sin² functions (lines).

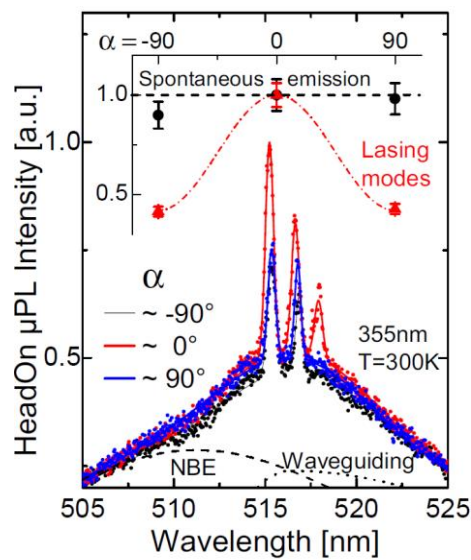

**Figure 2.** Nanowire amplified spontaneous emission (ASE) spectra measured in 'head-on' geometry at an optical pumping power around the threshold for the excitation polarization parallel ($\alpha = 0°$, red) and perpendicular ($\alpha = \pm 90°$, blue/black) to the nanowire axis show stimulated emission modes superimposed to dominating spontaneous emission, which consists of near band edge (NBE) and red-shifted waveguide emission. The inset shows the polarization dependence of the stimulated mode intensity (red triangles) and the nearly constant spontaneous emission (black circles).

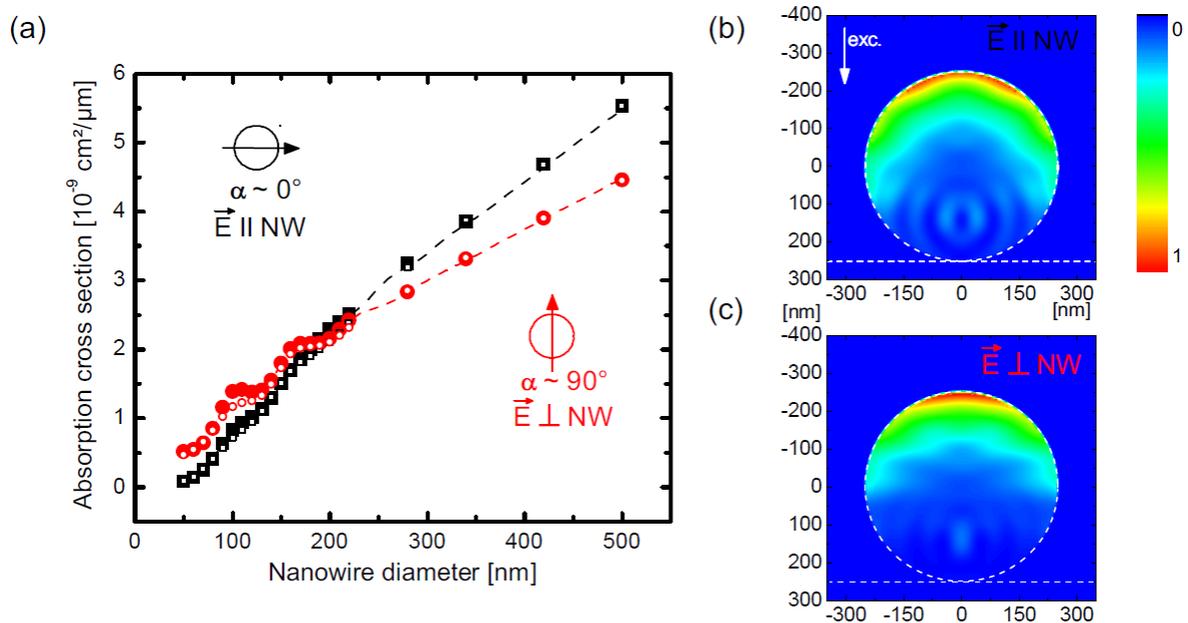

**Figure 3.** (a) Absorption cross section for parallel ($\alpha = 0°$, black) and perpendicular ($\alpha = 90°$, red) electric field respective to the nanowire axis for optical pumping as function of the diameter of the CdS nanowire completely surrounded by air (dashed line [analytical calculation][33] and filled symbols [FDTD simulation]) or lying on a SiO$_2$ substrate (open symbols [FDTD simulation]). Note that for diameters above 200nm the influence of the surrounding material is almost negligible and therefore filled and open symbols coincide. (b) and (c): Normalized absorption profiles (absorbed energy density normalized by incident intensity) for a CdS nanowire (diameter 500 nm) used in the experiment under parallel (b) and perpendicular (c) excitation polarization.

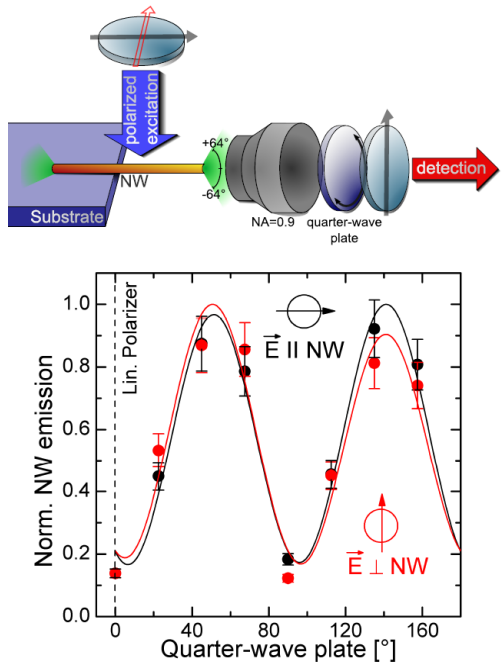

**Figure 4.** Modified 'head-on' measurement geometry to determine the Stokes parameter for both excitation polarizations (black: parallel, red: perpendicular to NW axis) using an objective (NA = 0.9), a quarter-wave plate and a linear polarizer [25]. The detected power as function of the quarter-wave plate position is afterwards used to determine the Stokes parameters of the NW laser emission.

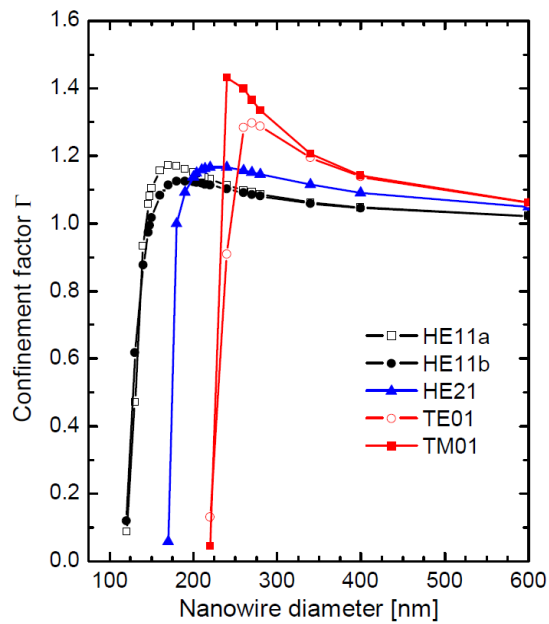

**Figure 5.** Confinement factors for the lowest order modes in a CdS nanowire on a SiO$_2$ substrate. The electric field of the HE11a mode is dominantly polarized in the direction pointing towards the substrate, the HE11b mode is mostly polarized longitudinally to the substrate surface. With increasing nanowire diameter the HE21 mode is very likely the first mode that starts to lase.

[1]     Piccione B, Cho C H, van Vugt L K and Agarwal R 2012 *Nat Nano* **7** 640–645

[2]     Moore G 1965 *Electronics* **38**

[3]     Agarwal R and Lieber C 2006 *Applied Physics A: Materials Science & Processing* **85** 209–215

[4]     Miller D 2009 *Proceedings of the IEEE* **97** 1166–1185

[5]     Kriesch A, Burgos S P, Ploss D, Pfeifer H, Atwater H A and Peschel U 2013 *Nano Letters* **13** 4539–4545

[6]     Landreman P E, Brongersma M L 2014 *Nano Letters* **14** 429–434

[7]     Bao J, Zimmler M A, Capasso F, Wang X and Ren Z F 2006 *Nano Letters* **6** 1719–1722

[8]     Zimmler M A, Voss T, Ronning C and Capasso F 2009 *Applied Physics Letters* **94** 241120

[9]     Hayden O, Greytak A and Bell D 2005 *Advanced Materials* **17** 701–704

[10]    Zimmler M, Bao J, Capasso F, Müller S and Ronning C 2008 *Applied Physics Letters* **93** 051101

[11]    Pan A, Liu D, Liu R, Wang F, Zhu X and Zou B 2005 *Small* **1** 980–983

[12]    Voss T, Svacha G, Mazur E, Müller S, Ronning C, Konjhodzic D and Marlow F 2007 *Nano letters* **7** 3675–3680

[13]    Qian F, Li Y, Gradecak S, Park H G, Dong Y, Ding Y, Wang Z L and Lieber C M 2008 *Nat Mater* **7** 701–706

[14]     Geburt S, Thielmann A, Röder R, Borschel C, McDonnell A, Kozlik M, Kühnel J, Sunter K A, Capasso F and Ronning C 2012 *Nanotechnology* **23** 365204

[15]     Saxena D, Mokkapati S, Parkinson P, Jiang N, Gao Q, Tan H H and Jagadish C 2013 *Nat Photon* **7** 963–968

[16]     Röder R, Wille M, Geburt S, Rensberg J, Zhang M, Lu J G, Capasso F, Buschlinger R, Peschel U and Ronning C 2013 *Nano Letters* **13** 3602–3606

[17]     Zimmler M, Capasso F, Müller S and Ronning C 2010 *Semiconductor Science and Technology* **25** 024001

[18]     Gao H, Fu A, Andrews S C and Yang P 2013 *Proceedings of the National Academy of Sciences* **110** 865–869

[19]     Xiao Y, Meng C, Wang P, Ye Y, Yu H, Wang S, Gu F, Dai L and Tong L 2011 *Nano letters* **11** 1122–1126

[20]     Agarwal R, Barrelet C and Lieber C 2005 *Nano letters* **5** 917–920

[21]     Giblin J, Protasenko V and Kuno M 2009 *ACS nano* **3** 1979–1987

[22]     Zhou R, Chang H, Protasenko V, Kuno M, Singh A, Jena D *et al.* 2007 *Journal of applied physics* **101** 073704–073704

[23]     Fan Z, Chang P, Lu J, Walter E, Penner R, Lin C and Lee H 2004 *Applied physics letters* **85** 6128

[24]     Taflove A and Hagness S C 2000 *Computational electrodynamics: The Finite-Difference Time-Domain Method* vol 160 (Artech House Boston)

[25]     Schaefer B, Collett E, Smyth R, Barrett D and Fraher B 2007 *American Journal of Physics* **75** 163–168

[26]     Bass M, DeCusatis C, Enoch J, Lakshminarayanan V, Li G, MacDonald C, Mahajan V and Van Stryland E 2009 *Handbook of Optics, Third Edition Volume IV: Optical Properties of Materials, Nonlinear Optics, Quantum Optics (set)* Handbook of Optics (McGraw-Hill Companies,Incorporated)

[27]     Versteegh M, Vanmaekelbergh D and Dijkhuis J 2012 *Physical Review Letters* **108** 157402

[28]     Klingshirn C 2005 *Semiconductor optics* vol 1439 (Springer Verlag Berlin Heidelberg)

[29]     Hurtado A, Xu H, Wright J B, Liu S, Li Q, Wang G T, Luk T S, Figiel J J, Cross K, Balakrishnan G, Lester L F and Brener I 2013 *Applied Physics Letters* **103** 251107

[30]     Maslov A, Ning C Z 2003 *Applied physics letters* **83** 1237

[31]     Maslov A, Ning C Z 2004 *Quantum Electronics, IEEE Journal of* **40** 1389–1397

[32]     Siegman A 1986 *Lasers* (University Science Books, Sausalito)

[33]     Bohren C F and Huffman D R 1998 *Absorption and scattering of light by small particles* (Wiley-Interscience, New York)

# Corrigendum: Polarization features of optically pumped CdS nanowire lasers (J. Phys. D: Appl. Phys. 47 (2014) 394012)

Robert Röder[1], Daniel Ploss[2], Arian Kriesch[2], Robert Buschlinger[2], Sebastian Geburt[1], Ulf Peschel[2] and Carsten Ronning[1]

[1] Friedrich-Schiller-Universität Jena, Institut für Festkörperphysik, 07743 Jena, Germany

[2] Friedrich-Alexander-Universität Erlangen-Nürnberg, Institut für Optik, Information und Photonik, and Erlangen Graduate School in Advanced Optical Materials (SAOT), Haberstr. 9a, 91058 Erlangen, Germany

Email: robert.roeder@uni-jena.de

In our recent publication [1] on the polarization feature of optically pumped CdS nanowire lasers, we labelled the transversal waveguide modes incorrectly due to a mistake in the visualisation of the FDTD simulation results. The HE21 mode turned out to be actually the TE01 mode, while the old TE01 (TM01) mode has now been identified as the new HE21a (HE21b) mode. Among the four lowest order modes HE11, TE01, TM01 and HE21, the TE01 mode exhibits the highest reflectivity of 56%, while the remaining modes have reflectivities in the range between 20% and 30%. The labels in figure 5 of ref. [1] need to be changed accordingly and will be replaced by figure 1. Thus, the diameter limit for low threshold lasing of 170-180 nm [14] coincides with the cut-off of the TE01 mode. The TE01 mode is therefore most likely the first mode starting to lase with increasing diameter, as the TE01 also requires the lowest material gain. In conclusion, the transversal lasing mode, which is most likely the TE01 mode, is still not notably affected by the polarization.

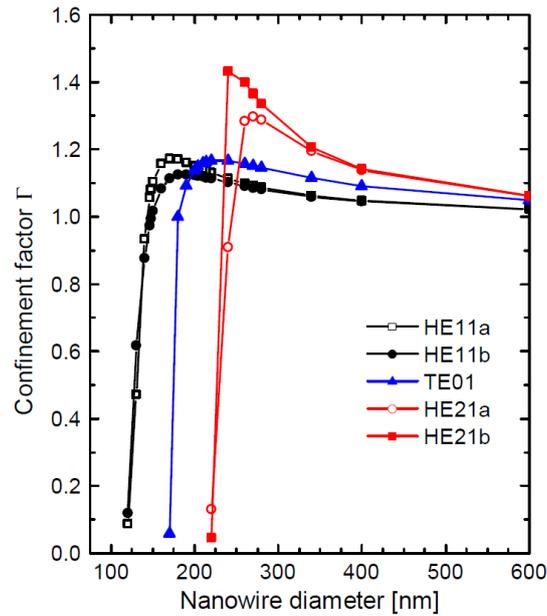

**Figure 1.** (Figure 5 in [1]) Confinement factors for the lowest order modes in a CdS nanowire on a $SiO_2$ substrate. The electric field of the HE11a mode is dominantly polarized in the direction pointing towards the substrate; the HE11b mode is mostly polarized longitudinally to the substrate surface. With increasing nanowire diameter the TE01 mode is very likely the first mode that starts to lase.

**References**

[1]     Röder R, Ploss D, Kriesch A, Buschlinger R, Geburt S, Peschel U and Ronning C 2014 *Journal of Physics D: Applied Physics* **47** 394012.

[2]     Geburt S, Thielmann A, Röder R, Borschel C, McDonnell A, Kozlik M, Kühnel J, Sunter K A, Capasso F and Ronning C 2012 *Nanotechnology* **23** 365204.